# MANIFOLD DAMPING OF TRANSVERSE WAKEFIELDS IN HIGH PHASE ADVANCE TRAVELING WAVE STRUCTURES AND LOCAL DAMPING OF DIPOLE WAKEFIELDS IN STANDING WAVE ACCELERATORS*

R.M. Jones*, SLAC; N.M. Kroll†, UCSD & SLAC; T. Higo, R.H. Miller*, and R.D Ruth*,
Stanford Linear Accelerator Center, Stanford, CA  94309

## Abstract

Operating the SLAC/KEK DDS (Damped Detuned Structure) X-band linacs at high gradients (in excess of 70MV/m) has recently been found to be limited by the accelerator structures breaking down and as a consequence severe damage occurs to the cells which makes the structures inoperable.  A series of recent experiments at SLAC indicates that arcing in the structures is significantly reduced if the group velocity of the accelerating mode is reduced and additionally it has been discovered that reducing the length of the accelerating structure also limits the number and intensity of breakdown events [1].  However, in designing new accelerating structures care must be taken to ensure that the beam-induced transverse wakefields do not cause the beam to become unstable.  Here, we report on damping transverse wakefields in two different short structures: a 90cm traveling wave structure in which the wakefield is coupled out to four attached manifolds and secondly, in a standing wave structure in which a limited number of cells heavily damp down the wakefield

*Supported under U.S. Department of Energy contract DE-AC03-76SF00515
†Supported under U.S. . Department of Energy contract DE-FG03-93ER40

# MANIFOLD DAMPING OF TRANSVERSE WAKEFIELDS IN HIGH PHASE ADVANCE TRAVELING WAVE STRUCTURES AND LOCAL DAMPING OF DIPOLE WAKEFIELDS IN STANDING WAVE ACCELERATORS *

R.M. Jones[*], SLAC; N.M. Kroll[†], SLAC; T. Higo, KEK; R.H. Miller*, and R.D Ruth*, SLAC

*Abstract*
Operating the SLAC/KEK DDS (Damped Detuned Structure) X-band linacs at high gradients (in excess of 70MV/m) has recently been found to be limited by the accelerator structures breaking down and as a consequence severe damage occurs to the cells which makes the structures inoperable. A series of recent experiments at SLAC indicates that arcing in the structures is significantly reduced if the group velocity of the accelerating mode is reduced and additionally it has been discovered that reducing the length of the accelerating structure also limits the number and intensity of breakdown events [1]. However, in designing new accelerating structures care must be taken to ensure that the beam-induced transverse wakefields do not cause the beam to become unstable. Here, we report on damping transverse wakefields in two different short structures: a 90cm traveling wave structure in which the wakefield is coupled out to four attached manifolds and secondly, in a standing wave structure in which a limited number of cells heavily damp down the wakefield

## 1. INTRODUCTION

The transverse wakefield is proportional to the transverse force on the beam integrated along the principal axis of the accelerator. The long-range transverse wakefield is an important parameter in the design of linear colliders because if left unimpeded it may cause transverse deflections of the beam which can lead to a dilution in the beam luminosity and to a BBU (Beam Break Up) instability. Over half a decade ago the transverse wakefield of the first 1.8 meter structure, of what would become a series, of damped and detuned structures (DDS) was measured at SLAC[2]. The wakefield was found to be well predicted by the circuit model and the built-in structure diagnostics were confirmed (namely, beam position monitoring and structure alignment through monitoring a small portion of the dipole signal radiated to the manifold). However, recently it has been found that these structures suffer from RF breakdown and this results in severe cell damage. In order to alleviate this problem shorter (90cm and 60cm) structures with a fundamental phase advance of $5\pi/6$ (rather than $2\pi/3$ as in the past RDDS series) have been designed at SLAC and KEK and these structures have been analysed with local damping in mind for future structures. It is the purpose of section 2 of this paper to investigate as to whether or not a manifold damped version of these LDS (locally damped structures) is able to damp the transverse wakefield adequately for this shorter structure at an accelerating phase advance of $5\pi/6$.

Also under consideration are standing wave structures of lengths 20cm and 30cm. These structures require a reduced incident energy to obtain accelerating gradients equivalent to the proposed high phase advance travelling wave structures and they have the added benefit that once a breakdown occurs at an iris then the field collapses and the power is immediately reflected with little power being dissipated directly in the structure itself [3]. In the third section a method for damping the transverse wakefield in a group of standing wave structures which taken together, allows a Gaussian fall off in the wake function. Additional damping is provided by heavily loading down a limited number of cells. The final calculation of the wakefield indicates that loading down 7 cells with a Q of 10 out of a total of 138 cells provides excellent damping of the lower dipole band.

## 2. MANIFOLD DAMPED WAKES IN TRAVELING-WAVE STRUCTURES

In order to calculate the beam-induced wakefield in a multi-cell structure we utilize a circuit model to calculate the dispersion curves of the structure. The circuit model described in [4] is modified to take into account variable cell loading through resistors placed in series with each capacitor. We focus on the structure under design at SLAC

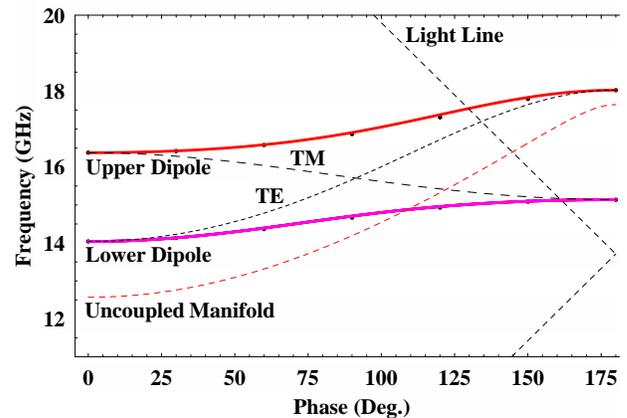

Figure 1: Dispersion curve for cell 42 of detuned accelerator (prior to manifold loading). The manifold curve prior to coupling to the main accelerator structure is shown dashed in red.

*Supported under U.S. DOE contract DE-AC03-76SF00515
†Supported under U.S. DOE grant DE-FG03-93ER40

namely, H90VG5, a structure consisting of 83 cells in which the initial group velocity of the accelerating mode is 0.0506% of the velocity of light and the phase advance at the accelerating frequency (11.424 GHz.) is chosen to be $5\pi/6$. This structure has been designed with a 2-dimensional computer code Omega2D [5] without attaching manifolds. In order to model the effect of adding four manifolds the dipole dispersion curves are supplemented by the manifold dispersion curve (prior to coupling to the structure). The parameters for the manifold are chosen to be similar to those used in the design RDDS1 [6] in order that the coupling to the manifold and the effective susceptance of the iris coupling to the manifold remain similar and physically realizable quantities. In this design process we chose three representative cells (1,42 and 83) and from these cells we obtain all remaining parameters by fitting quadratic functions to the known parameters as function of the synchronous frequency. The Brillioun diagram for cell 42 is shown in Fig. 1 and after coupling to the manifold the new dispersion curves are shown in Fig 2.

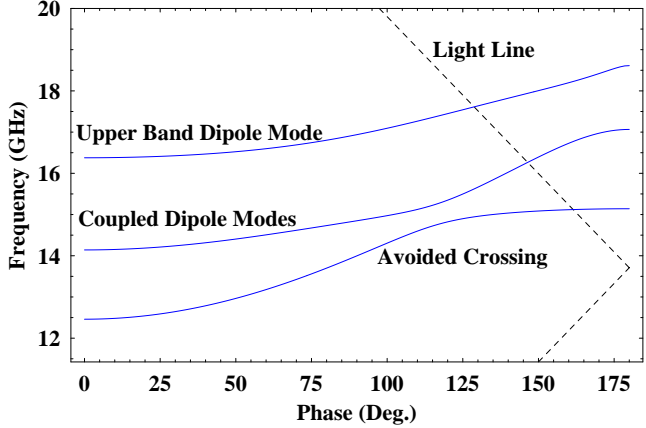

Figure 2: Dispersion curve for cell 42 of DDS

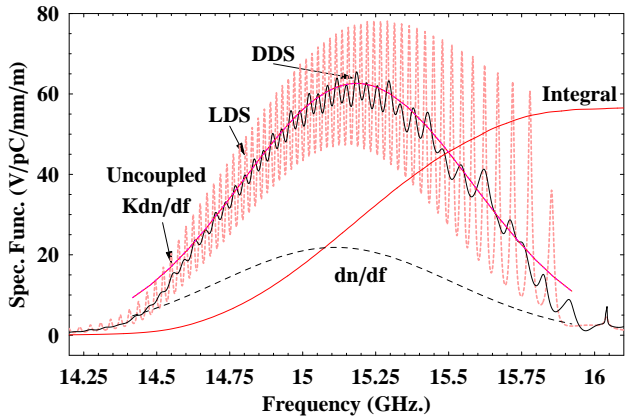

Figure 3: Spectral function for damped detuned structure (DDS) compared to a locally damped stucture in which each individual cell has a Q of 1000.

The lower two curves are now mixed modes between the manifold and the dipole mode of the accelerating structure and they avoid crossing each other due to the coupling. The intensity of the coupling of the dipole mode to the manifold mode is directly proportional to the minimum separation of the lower two dispersion curves.

The wake function is calculated by utilizing the spectral function method [7] and this technique is applied to the LDS (locally damped structure) and to the same structure in which a manifold is coupled through coupling slots at each cell. The spectral functions that result from this calculation are shown in Fig. 3 and it is evident that the DDS shows superior damping properties compared to the LDS

The inverse transform of the spectral function enables the wake function to be calculated and this is shown in Fig. 4. The wake field is well damped in the DDS compared to the LDS where BBU will occur unless the structure frequencies are interleaved [8].

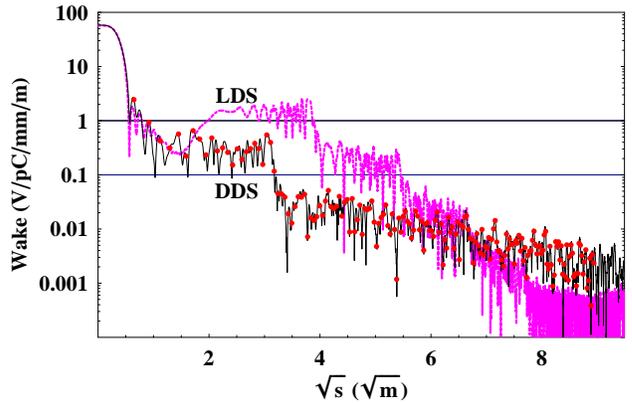

Figure 4: Envelope of wake function for the manifold damped structure. The distance behind the driving bunch is given by s.

## 3. DAMPING WAKEFIELDS IN STANDING WAVE ACCELERATOR STRUCTURES

To demonstrate the damping of dipole wakefields in standing wave structures we consider an amalgamated structure consisting of 6 individual accelerating structures each of which has 23 cells (making a total of 138 cells).

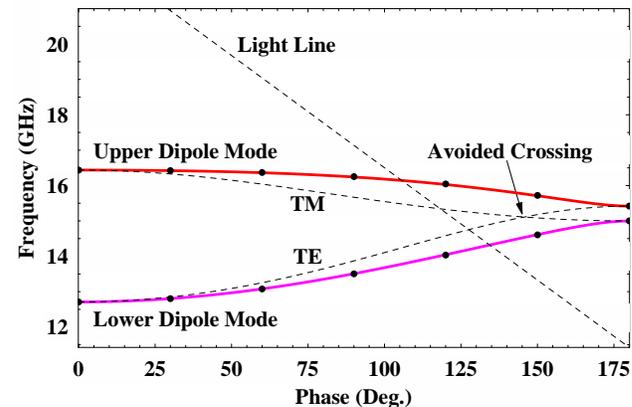

Figure 5: Dipole dispersion curve for cell 69 of an amalgamated standing wave structure.

The mode density of the "super-structure" is made to follow a Gaussian functional form by enforcing an error function variation on the tapering down of the iris dimensions with increasing cell number. This makes the wakefield initially fall off rapidly with a Gaussian distribution but as there are of course a finite number of cells then the wake recoheres and thus additional damping is required. We consider direct local damping to accomplish the task of maintaining the fall off in the wakefield  The local damping is limited to the first cell of each individual structure so that there are in total only 7 cells which we load down heavily with a Q of 10 for both the TE and TM modes. We envisage the cells in which the dipole mode are damped to be cut-off to the accelerating mode and thus to have little impact on the shunt impedance of that mode. The practicality of simultaneously loading down both TE and TM modes is currently under investigation.

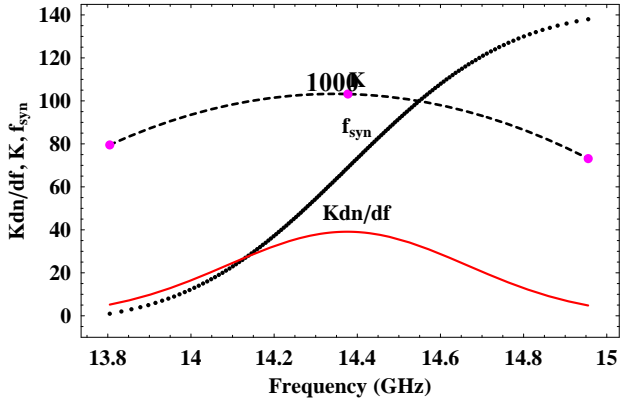

Figure 6: Uncoupled kick Factors (K), synchronous frequency distribution ($f_{syn}$) and kick factor weighted density function (Kdn/df) for standing wave structure.

Three cells are chosen to parameterize the dipole characteristics of the structure and the dispersion curve for cell 69 is shown in Fig. 5. The zero and $\pi$ modes are obtained with the computer code Omega2D[5] and the efficacy of the circuit model is attested to by how well the remaining points fit the curves generated. The synchronous point (shown in Fig. 5 and Fig. 6) is considerably shifted away from the $\pi$ point compared to the traveling wave structure and thus the lower band TE modal content will be significantly enhanced. The lower band kick factors (K, shown in Fig. 6) do not have a linear variation with synchronous frequency as they did in all previous traveling wave structures designed at SLAC even though the TM content increases as the irises pinch down with increasing cell number. However, the TE content also increases markedly and this leads to a decrease in the kick factors of the first band towards the end of the amalgamated structure. This of course has a beneficial effect on the wakefield as the spectral function (shown in Fig. 7) has an almost equal spacing of modes in the upper and lower frequency ends (in contra-distinction to all traveling wave structures previously considered in which the high frequency modes were widely spaced relative to the lower band modes). The wake function (shown in Fig. 8), obtained as a Fourier transform of the spectral function benefits from this new modal distribution in that the damping of the wakefield is well sustained. The wake initially damps with a Gaussian roll-off due to the detuning of the cells and at later distances the loaded Q of the modes starts to play a role. In the region 0.42cm to 400cm the Q is approximately 300.

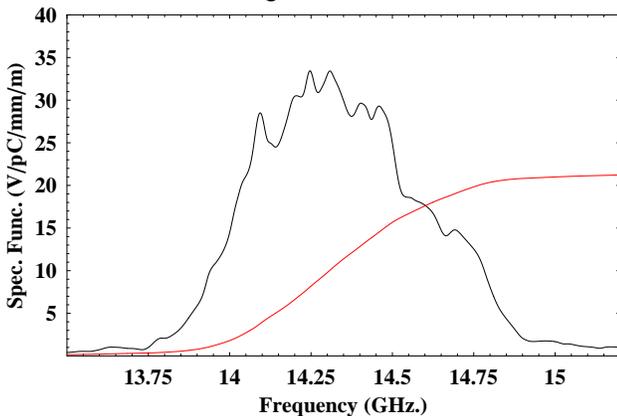

Figure 7: Spectral function of the lower band frequencies of a standing wave structure consisting of 6 separate structures (each structure containing 23 cells).

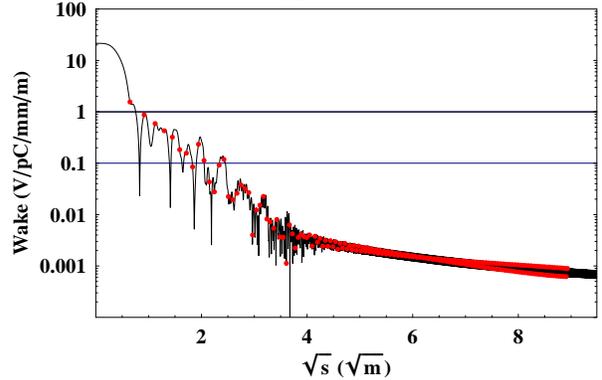

Figure 8: Envelope of wake function for a 138-cell standing wave amalgamated structure.

However, higher order dipole bands must also be analyzed. Preliminary calculations on the second band, for example, indicate that the kick factors have a similar magnitude as the lower dipole band and these will also require damping.

## 4. REFERENCES

[1] C. Adolphsen, ROAA003, this conf.
[2] R.H. Miller et al, LINAC96, also SLAC-PUB 7288
[3] R.D. Ruth, HEPAP, SLAC, 2001
[4] R.M. Jones et al, EPAC96, also SLAC-PUB 7187
[5] X. Zhan, Ph.D. thesis, Stanford University, 1997
[6] R.M. Jones et al, PAC99, also SLAC-PUB 8103
[7] R.M. Jones et al, LINAC96, also SLAC-PUB 7286
[8] R.M. Jones et al, MPPH058, this conf.